\documentclass[letter]{aa}

\usepackage{amsmath,amssymb}
\usepackage{graphicx}
\usepackage{natbib}
\usepackage{txfonts}
\usepackage{xcolor}
\usepackage[colorlinks=true,linkcolor=blue,citecolor=magenta,urlcolor=cyan]{hyperref}

\bibpunct{(}{)}{;}{a}{}{,}
\graphicspath{{./}}

\newcommand{\Lbol}{L_{\rm bol}}
\newcommand{\Rbol}{R_{\rm bol}}

\newcommand{\tff}{\tau_{\rm ff}}
\newcommand{\ergs}{\,{\rm erg\,s^{-1}}}
\newcommand{\ergshz}{\,{\rm erg\,s^{-1}\,Hz^{-1}}}
\newcommand{\Ne}{N_e}
\renewcommand{\ne}{n_e}

\title{Radio constraints on obscuring columns in local little red dots}
\titlerunning{Radio constraints on obscuring columns in local LRD analogues}
\authorrunning{An et al.}

\author{Tao An\inst{1}\thanks{Corresponding author: antao2008@ustc.edu.cn}
\and Ailing Wang \inst{2} \and Ruqiu Lin\inst{3} }
\institute{Department of Astronomy, University of Science and Technology of China, Hefei, Anhui 230026, China
\and Yunnan Observatories, Chinese Academy of Sciences, Kunming 650216, China \and Department of Astronomy, University of Massachusetts Amherst, Amherst, MA 01003, USA }

\date{Received ; accepted }

\begin{document}

\abstract
%Context
{}
%Aims
{The ``little red dots'' (LRDs) discovered by the James Webb Space Telescope (JWST) are modelled as accreting black holes embedded in compact ionised gas with electron column densities $\Ne\sim10^{24}\,{\rm cm^{-2}}$. We test whether the radio components detected in local LRD analogues can propagate through the full electron column along the line of sight.}
%Methods
{We derive a free--free opacity bound that holds for any density distribution at fixed electron column $N_e$, path length $\Delta R$, and maximum electron temperature $T_e$. We compare the emission measure required by the scattering column with the maximum allowed by the 144\,MHz detections of the V-shaped LRD analogues J0923 and J1200. The radio spectrum of J1047 and the radio properties of 44 local LRD analogues provide supporting tests.}
% Results
{For $T_e=10^4$\,K and $\Delta R=0.3$\,pc, free--free opacity $\tff=1$ is reached at $\Ne\simeq10^{22}\,{\rm cm^{-2}}$ at rest-frame 3\,GHz and at only $\simeq5\times10^{20}\,{\rm cm^{-2}}$ at the rest-frame 190\,MHz of the J0923 and J1200 detections. Even for an intrinsic radio power $2\times10^4$ times the observed one, a $10^{24}\,{\rm cm^{-2}}$ column requires an emission measure over two orders of magnitude above that allowed by the detections, for any $T_e\leq10^6$\,K and $\Delta R\leq1$\,pc. The turnover of J1047 can be explained either by free--free absorption in gas with a much lower emission measure or by synchrotron self-absorption. The V-shaped and non-V-shaped comparison samples show no significant difference in radio detection fraction.}
%Conclusion
{If the radio-detected LRDs such as J0923 and J1200 carry scattering columns as large as those adopted in dense-gas models, their detected radio emission must originate outside the scattering layer or reach the observer along paths of much smaller electron column. The detections thus constrain the foreground electron columns along the paths to the radio components.}

\keywords{galaxies: active -- galaxies: nuclei -- radio continuum: galaxies -- galaxies: high-redshift}

\maketitle
\nolinenumbers

%===================================================================
\section{Introduction}
\label{sec:intro}
%===================================================================

The James Webb Space Telescope (JWST) has revealed an abundant population of compact red sources at $z>4$ \citep[e.g.,][]{2024ApJ...963..129M,2024ApJ...964...39G}, known as ``little red dots'' (LRDs), whose physical nature remains debated. A proposed unifying framework proposes that an accreting black hole is embedded in a compact, sub-pc ionised cocoon with an electron column density of $\Ne\sim10^{24}\,{\rm cm^{-2}}$ \citep{2026Natur.649..574R,2026arXiv260118864S}. 
In this picture, electron scattering produces the exponential wings of the broad emission lines, while the dense gas reprocesses the nuclear radiation to produce the red optical continuum, either through recombination emission
\citep{2026arXiv260118864S} or through emission from a photosphere with $T\sim4000$--$6000$\,K \citep{2026NatAs..10.1261I}.
These diagnostics probe the inner gas at optical, near-infrared, and X-ray wavelengths. Radio observations add a complementary constraint. \citet{2026aask.bookE..77M} quantified the free--free attenuation of active galactic nuclei (AGN) radio emission by the ionised scattering layer and predicted the detectability of the attenuated emission with the Square Kilometre Array (SKA).  \citet{2026arXiv260920913T} showed how the spectral turnover of thermal free--free emission of the surrounding gas constrains its column density and spatial size. These studies motivate a direct observational test of whether the radio emission already detected in local LRD analogues can propagate through the full electron columns invoked by dense-cocoon models. This test constrains the geometry of the radio-emitting
regions relative to the obscuring gas.

Several groups have identified low-redshift ($z<1$) galaxies that reproduce many defining LRD signatures, including broad Balmer lines, V-shaped SEDs, extreme Balmer decrements, Balmer and He\,{\sc i} absorption, and X-ray weakness \citep{2025ApJ...980L..34L,2026ApJ...997..364L,2026arXiv260521574L,2026A&A...709L..11D,2026ApJ...999...30C,2026arXiv260514233P}. These systems offer a nearby laboratory for testing the dense-cocoon picture at radio wavelengths. Paper I of this series \citep{An2026a} established the radio counterpart and long-term variability of J0923, while Paper II \citep{Wang2026} presented a radio census of 44 local LRD analogues with diverse radio properties. Here we apply a conservative free--free opacity bound at fixed electron column, path length, and maximum temperature that holds for any density distribution, and compare the emission measure required by the scattering column with that allowed by the radio detections. The 144\,MHz detections of J0923 and J1200 provide the most stringent constraints (Sect.~\ref{sec:joint}). We also examine J1047 (Sect.~\ref{sec:j1047}) and the 44 analogues (Sect.~\ref{sec:efficiency}).

%===================================================================
\section{A conservative free--free opacity bound}
\label{sec:opacity}
%===================================================================

The free--free optical depth of ionised gas scales as $\tff\propto T_e^{-1.35}\nu^{-2.1}\,{\rm EM}$, with emission measure ${\rm EM}=\int\ne^2\,dl$ (Eq.~\ref{eq:tff}, Appendix~\ref{app:freefree}). 
For a line-of-sight path of length $\Delta R$ with electron column density $\Ne=\int\ne\,dl$, the Cauchy--Schwarz inequality gives ${\rm EM}\geq\Ne^2/\Delta R$, with equality
for a uniform-density slab.
Clumping therefore only raises $\tff$, and a temperature gradient cannot lower it below that of a uniform screen at the maximum $T_e$ (Appendix~\ref{app:freefree_cs}). \citet{2026aask.bookE..77M} computed the free--free attenuation of radio emission by the scattering layer. We cast it as a conservative bound that applies to individual detected components for any density structure along the path.

\begin{figure}
\centering
\includegraphics[width=\columnwidth]{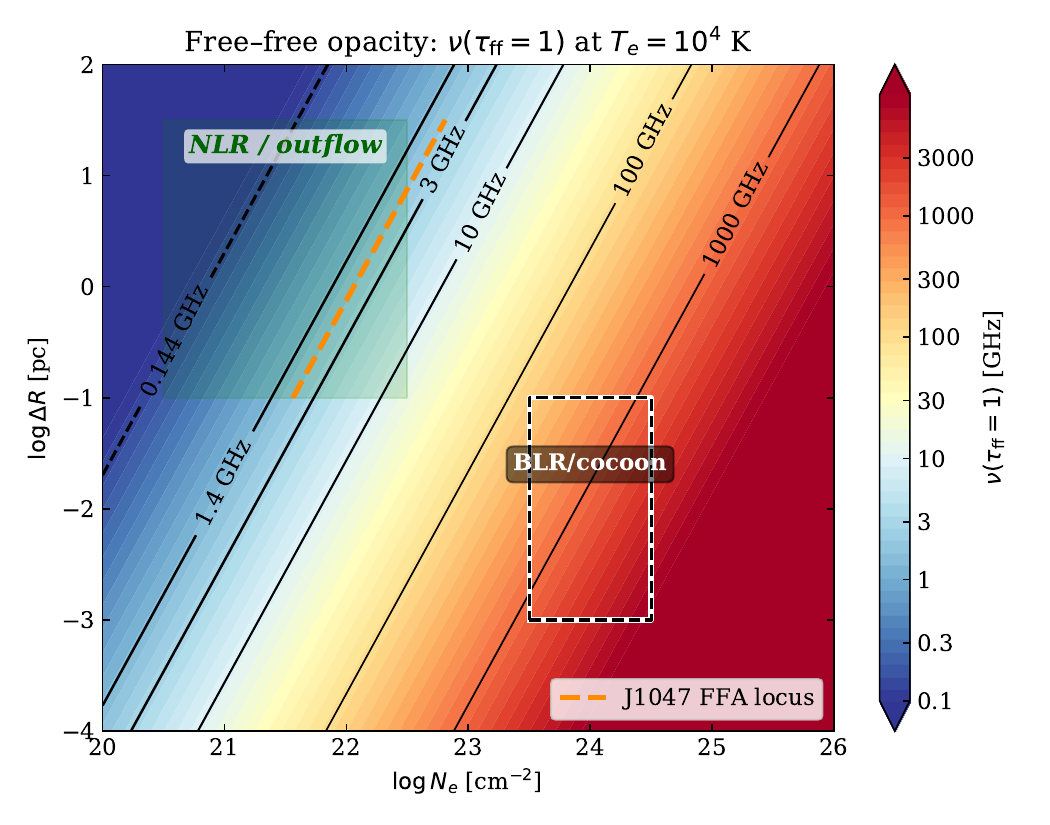}
\caption{Free--free optical depth $\tff=1$ frequency as a function of electron column $\Ne$ and absorber thickness $\Delta R$, at $T_e=10^4$\,K. The dashed box marks the cocoon parameter space ($\Ne\sim10^{23.5}$--$10^{24.5}\,{\rm cm^{-2}}$, $\Delta R\sim10^{-3}$--$10^{-1}$\,pc) from broad-line modelling \citep{2026Natur.649..574R}. The green box marks typical pc-scale narrow-line-region (NLR) and outflow parameters. The orange dashed line is the locus of constant EM fitted to J1047 with the homogeneous free--free absorption (FFA) model, and its extent in $\Delta R$ is illustrative.}
\label{fig:opacity}
\end{figure}

Cocoon models infer $\Ne\sim10^{24}\,{\rm cm^{-2}}$ from the scattering wings and a scattering region of a few light days (${\sim}10^{16}$\,cm) at the adopted densities \citep{2026Natur.649..574R,2026arXiv260118864S}. For $T_e=10^4$\,K and a representative $\Delta R=3\times10^{-3}$\,pc, this column gives $\tff(3\,{\rm GHz})\sim10^{6}$, and $\tff=1$ is reached only at $\nu\sim2300$\,GHz. A $1$\,pc path, ten times the compact-cocoon range of Fig.~\ref{fig:opacity}, still gives $\tff(3\,{\rm GHz})\sim3\times10^{3}$, and $T_e=10^6$\,K with $\Delta R\lesssim0.1$\,pc gives $\tff(3\,{\rm GHz})\gtrsim68$ (Table~\ref{tab:robust}). A sightline through the full scattering column is therefore opaque at centimetre wavelengths.

Setting $\tff=1$ defines the critical electron column density separating the optically thin and thick regimes,
\begin{equation}
\Ne^{\rm crit} =
1.7\times10^{22}
\left(\frac{\nu}{3\,{\rm GHz}}\right)^{1.05}
\left(\frac{\Delta R}{1\,{\rm pc}}\right)^{0.5}
\left(\frac{T_e}{10^4\,{\rm K}}\right)^{0.675}
\,{\rm cm^{-2}},
\label{eq:Ne_limit}
\end{equation}
with $\nu$ measured in the absorber frame. For $T_e=10^4$\,K and $\Delta R=0.3$\,pc, larger than the assumed compact-cocoon scale, $\Ne^{\rm crit}\simeq10^{22}\,{\rm cm^{-2}}$ at 3\,GHz, two orders of magnitude below the cocoon column. Detectable emission can still emerge at $\tff>1$ when the intrinsic source is bright. For a foreground screen covering the component and an upper limit $S_{\rm int,max}$ on the intrinsic flux density, $\tff\leq\tau_{\max}=\ln(S_{\rm int,max}/S_{\rm obs})$, which raises the column limit to $\Ne^{\rm crit}\tau_{\max}^{1/2}$. For J0923 and J1200, detected at observed 144\,MHz (rest frame ${\simeq}190$\,MHz), $\Ne^{\rm crit}$ is 18 times lower than at 3\,GHz, corresponding to $5\times10^{20}\,{\rm cm^{-2}}$.

%===================================================================
\section{Emission measure constraints for J0923 and J1200}
\label{sec:joint}
%===================================================================

\noindent
J0923 ($z=0.323$) and J1200 ($z=0.303$) exhibit the V-shaped continuum, a key feature motivating dense-gas interpretation of LRDs \citep{2025ApJ...980L..34L}. Both are securely detected by the LOFAR Two-metre Sky Survey (LoTSS, \citealt{2022A&A...659A...1S}) at observed 144\,MHz, or rest-frame $\simeq 190$\,MHz. Their radio luminosities exceed those expected from the standard AGN coronal radio--X-ray scaling by factors of 17--53, assuming X-ray luminosities inferred from $L_{\rm bol}$ (Appendix~\ref{app:corona}). We compare two emission measures for a common foreground path. For a column $\Ne$ distributed in a line-of-sight path of length $\Delta R$, Eq.~(\ref{eq:cs}) gives the minimum emission measure implied by the scattering column, regardless of the density distribution
\begin{equation}
{\rm EM}_{\rm min,opt}=1.05\times10^{11}
\left(\frac{\Ne}{10^{24}\,{\rm cm^{-2}}}\right)^{2}
\left(\frac{\Delta R}{1\,{\rm pc}}\right)^{-1}
{\rm pc\,cm^{-6}}.
\label{eq:em_opt}
\end{equation}
A foreground screen covering the component, with optical depth $\tau_{\max}$ at rest frequency $\nu$, gives the radio upper bound, from Eq.~(\ref{eq:tff}),
\begin{equation}
{\rm EM}_{\rm max,radio}=3.05\times10^{6}\,\tau_{\max}
\left(\frac{T_e}{10^4\,{\rm K}}\right)^{1.35}
\left(\frac{\nu}{\rm GHz}\right)^{2.1}
{\rm pc\,cm^{-6}}.
\label{eq:em_radio}
\end{equation}
We adopt $\tau_{\max}=10$ as a generous attenuation budget at the detected frequency, allowing an intrinsic flux density $e^{10}\simeq2\times10^{4}$ times the observed value. Since $Q\propto\tau_{\max}^{-1}$ (see below), doubling this budget still leaves $Q\gtrsim100$ for $\Ne=10^{24}\,{\rm cm^{-2}}$, $T_e\leq10^6$\,K, and $\Delta R\leq1$\,pc. At $\nu=0.19$\,GHz, ${\rm EM}_{\rm max,radio}\simeq9\times10^{5}(T_e/10^4\,{\rm K})^{1.35}\,{\rm pc\,cm^{-6}}$.

We define the ratio $Q\equiv{\rm EM}_{\rm min,opt}/{\rm EM}_{\rm max,radio}=(\Ne/\Ne^{\rm crit})^2/\tau_{\max}$ to assess whether the two constraints are compatible with a common foreground column. 
A value of $Q>1$ rules out the corresponding parameter combination for a fully covering foreground screen.
For $\Ne=10^{24}\,{\rm cm^{-2}}$ and $T_e=10^4$\,K, $Q\simeq10^{5}$--$10^{6}$ for $\Delta R=0.1$--$1$\,pc, path lengths already beyond the compact-cocoon scale, and $Q\simeq200$--$2000$ for $T_e=10^6$\,K. 
Reducing the electron column density to $10^{23.5}\,{\rm cm^{-2}}$ still gives $Q\gtrsim20$ for $T_e\leq10^6$\,K and $\Delta R\leq1$\,pc. Gas at $T_e\sim10^7$\,K spread over $\Delta R\gtrsim1$\,pc with $\Ne\lesssim10^{23.5}\,{\rm cm^{-2}}$ can approach $Q\sim1$, a configuration far from the cool, light-day-scale layers of current models. Density structure and temperature gradients only raise these values (Appendix~\ref{app:freefree_cs}).

The test adopts the scattering column of dense-gas models for both objects, treats the absorber as a foreground screen that covers the detected component, and assumes the observed 144\,MHz emission originates near the nucleus. Under these premises, the detections exclude a radio component behind the full compact, cool scattering column. Emission from outside an otherwise spherical layer, including extended star formation or jet emission within the unresolved LoTSS beam, from a low-column surface layer, or through low-column channels remains allowed (Appendix~\ref{app:freefree_paths}).

%===================================================================
\section{Low-frequency suppression of J1047}
\label{sec:j1047}
%===================================================================

\begin{figure}
\centering
\includegraphics[width=0.9\columnwidth]{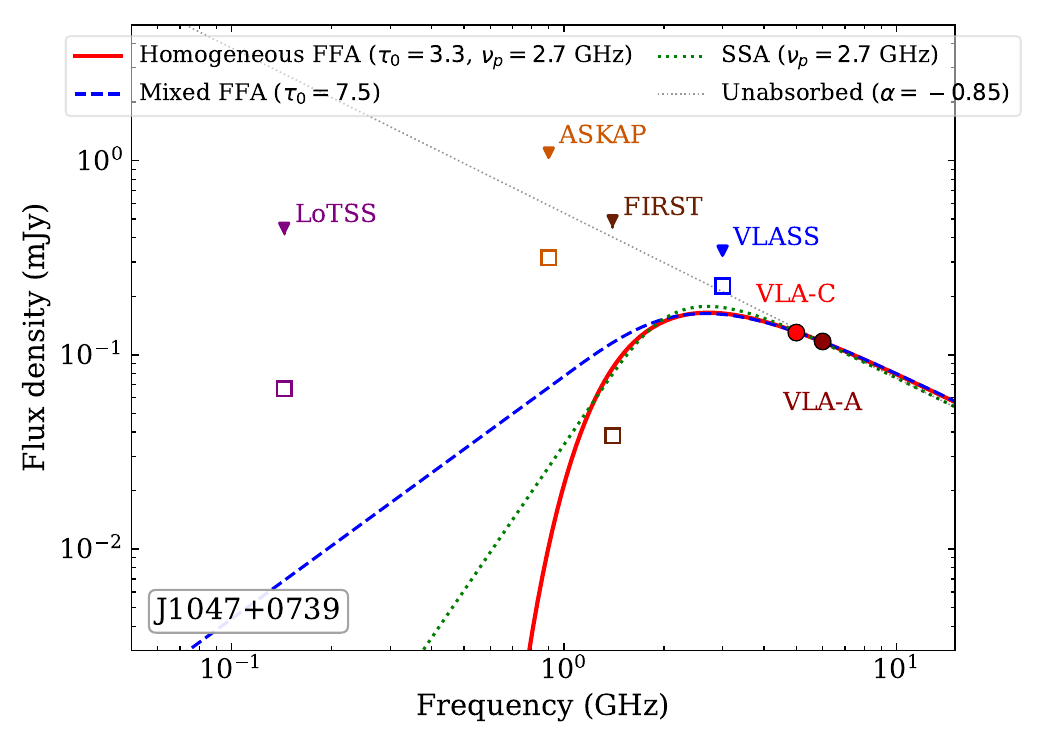}
\caption{Radio spectrum of J1047+0739. Filled circles at 5 and 6\,GHz are VLA detections from \citet{2026A&A...707L..17R}. Open squares at 144\,MHz, 0.9, and 1.4\,GHz are forced-photometry non-detections from Paper~II \citep{Wang2026}, with downward arrows denoting the corresponding $3\sigma$ upper limits. The open diamond at 3\,GHz shows the VLASS three-epoch mean, below the detection threshold ($0.23\pm0.10$\,mJy, $2.3\sigma$). The solid curve shows a homogeneous FFA model with $\alpha_{\rm thin}=-0.85$ and $\tau_0(1\,{\rm GHz})=3.3$, the dashed curve a mixed FFA model with $\tau_0=7.5$, and the dotted curve an SSA model peaking at $\nu_p\simeq2.7$\,GHz.}
\label{fig:j1047}
\end{figure}

J1047+0739 ($z=0.168$, hereafter J1047) shows the strongest low-frequency suppression in the sample presented in Paper II \citep{Wang2026}. It is detected at 6\,GHz with $S_6=117\pm8\,\mu$Jy and a spectral index $\alpha=-0.85\pm0.24$ across the 4--8\,GHz band \citep{2026A&A...707L..17R}, but remains undetected by LoTSS at 144\,MHz ($<0.39$\,mJy, $3\sigma$) \citep{Wang2026}. Extrapolating the 6\,GHz spectrum predicts $S_{144}\simeq2.8$\,mJy, seven times above the LoTSS limit. A slope flatter by $1\sigma$ ($\alpha=-0.61$) predicts 1.1\,mJy, still 2.8 times above the limit, and one flatter by $2\sigma$ approaches the limit. The three-epoch mean of the Very Large Array Sky Survey (VLASS, \citealt{2020PASP..132c5001L}), $0.23\pm0.10$\,mJy at 3\,GHz ($2.3\sigma$), agrees with the optically thin extrapolation of $0.21$\,mJy. The spectrum, assembled from non-simultaneous data, therefore favours low-frequency suppression, with illustrative fits peaking near 2.7\,GHz (Fig.~\ref{fig:j1047}). For fixed $\alpha_{\rm thin}$ the 95\% range of the peak is 0.8--5.8\,GHz (Appendix~\ref{app:j1047_models}).

Free--free absorption (FFA) fits with a homogeneous foreground screen and with absorbing gas mixed into the emitting region give respectively $\tau_0=3.3$ and $7.5$ at observed 1\,GHz (Appendix~\ref{app:j1047_models}), or rest-frame ${\rm EM}\simeq(1.4$--$3.2)\times10^7\,{\rm pc\,cm^{-6}}$ for $T_e=10^4$\,K (EM scales as $T_e^{1.35}$). The spectrum constrains the EM and leaves the absorber thickness free. 
For $\Delta R=0.1$--$1$\,pc, a uniform slab has $\Ne\sim(4$--$17)\times10^{21}\,{\rm cm^{-2}}$, an upper limit for any density structure and about two orders of magnitude below the scattering column, and a thinner absorber with $\Delta R=3\times10^{-3}$\,pc needs only $6\times10^{20}\,{\rm cm^{-2}}$. An FFA interpretation therefore implies an emission measure far below the $\gtrsim10^{11}\,{\rm pc\,cm^{-6}}$ of a compact $10^{24}\,{\rm cm^{-2}}$ column (Eq.~\ref{eq:em_opt}), regardless of the absorber's thickness or location. Synchrotron self-absorption (SSA) provides an alternative fit with a peak near 2.7\,GHz (Appendix~\ref{app:j1047_ssa}). A source near equipartition ($T_b\simeq5\times10^{10}$\,K) without Doppler boosting would have $\theta\simeq0.027$\,mas, corresponding to a projected size of 0.08\,pc, much larger than the light-day scattering region, while a higher brightness temperature allows a smaller source. In both models, the 5--6\,GHz detection already excludes a path through the full compact, cool scattering column, for which $\tff(6\,{\rm GHz})\gg1$ (Sect.~\ref{sec:opacity}). For $T_e=10^6$\,K and $\Delta R=1$\,pc, a $10^{24}\,{\rm cm^{-2}}$ column gives only $\tff\simeq1$ at the rest frequency of 6\,GHz. Simultaneous broadband spectra from 0.3 to 8\,GHz at matched resolution, together with very long baseline interferometry (VLBI) constraints on structure and brightness temperature, can help distinguish FFA from SSA.

%===================================================================
\section{Population context of 44 local LRD analogues}
\label{sec:efficiency}
%===================================================================

\begin{figure}
\centering
\includegraphics[width=0.9\columnwidth]{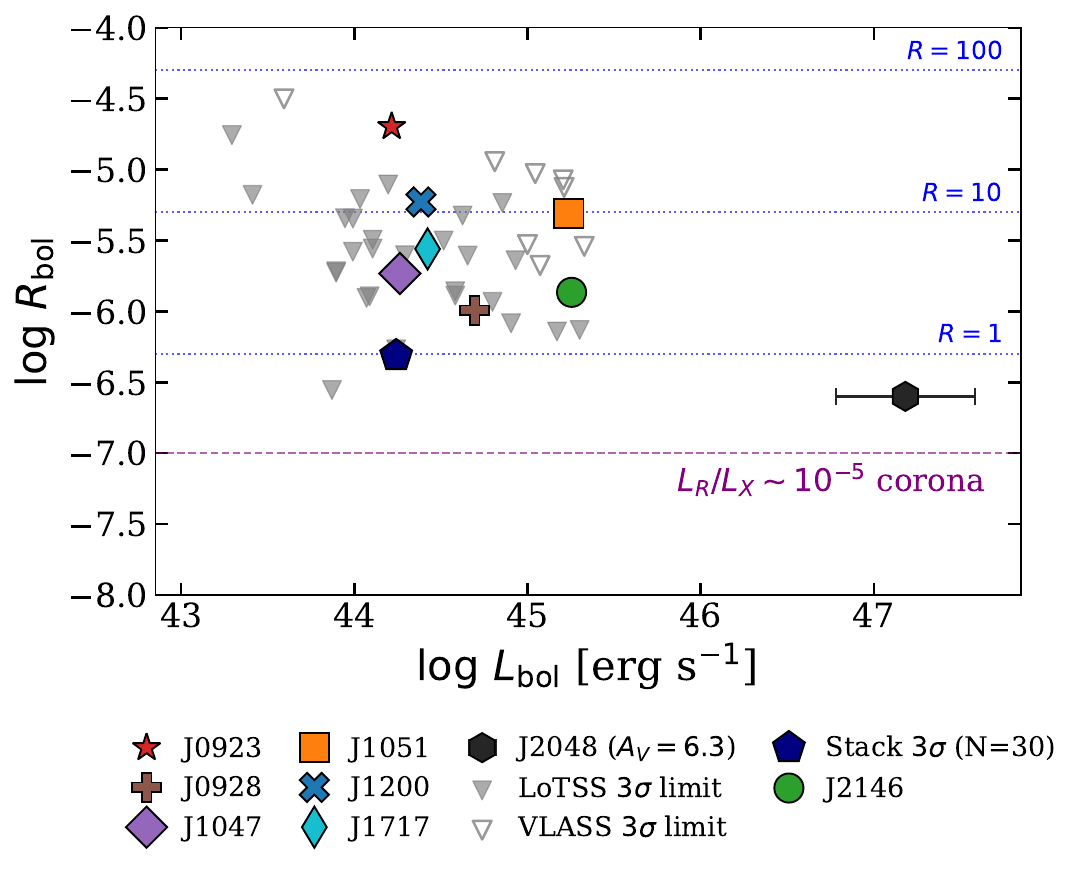}
\caption{Radio-to-bolometric luminosity ratio $\Rbol=\nu L_\nu(1.4\,{\rm GHz})/\Lbol$ versus bolometric luminosity for the 43 local LRD analogues with available $\Lbol$. Filled symbols mark the eight radio detections. Grey triangles mark $3\sigma$ upper limits from LoTSS (filled) or VLASS (open). Dotted lines mark Kellermann ratios $\mathcal{R}=1$, 10, and 100 ($\log\Rbol=-6.3$, $-5.3$, $-4.3$) for $\Lbol=10\,L_B$ and $\alpha=-0.7$. The pentagon marks the LoTSS stacking limit at the median $\log\Lbol\simeq44.2$ and $z=0.43$ of the 25 stacked sources. J2048 is shown at the extinction-corrected luminosity of \citet{2026ApJ...999...30C}, with the $\pm0.4$\,dex uncertainty as a bar.}
\label{fig:Rbol}
\end{figure}

\noindent
The 44 local LRD analogues are drawn from six parent catalogues (Appendix~\ref{app:sample}). Papers II and III use the same  catalogue and signed radio measurements.
% (common_master_44.csv, shared-forced-20261011). 
% Their LoTSS 144\,MHz and VLASS 3\,GHz flux densities are taken from the forced-photometry catalogue of Paper~II \citep{Wang2026}. 
We use the radio-to-bolometric luminosity ratio $\Rbol=\nu L_\nu(1.4\,{\rm GHz})/\Lbol$ on the homogenised $\Lbol$ scale of Appendix~\ref{app:lbol}. Eight sources have a radio detection in LoTSS, VLASS, or the published VLA data (Fig.~\ref{fig:Rbol}). The nominal $3\sigma$ VLASS group contains six sources. For $\Lbol=10\,L_B$ and $\alpha=-0.7$ \citep{2007ApJ...658..815S},  we obtain  $\log\Rbol\simeq\log\mathcal{R}-6.3$, where $\mathcal{R}$ is the radio loudness parameter \citep{1989AJ.....98.1195K}.  J0923 and J1200 have $\mathcal{R}\simeq40$ and 13, comparable to radio-detected narrow-line Seyfert 1 galaxies. The inverse-variance weighted means of the 30 LoTSS and 38 VLASS non-detections, $0.008\pm0.015$\,mJy and $0.017\pm0.014$\,mJy, show no significant stacked signal (Appendix~\ref{app:stacking}).

\label{sec:control}The V-shaped and non-V-shaped broad-line Green Pea galaxies (BLGPs) of \citet{2025ApJ...980L..34L} offer a comparison drawn from the same parent population (Appendix~\ref{app:control}). Secure detections (${\rm S/N}\geq5$) give $4/11$ (non-V-shaped) versus $3/7$ (V-shaped) at 144\,MHz and $4/12$ versus $1/7$ at 3\,GHz. Two-sided Fisher exact tests of these fractions and of those including tentative detections (${\rm S/N}\geq3$) give $p=0.33$--$1.0$. Therefore the two classes show no statistically  significant difference in radio detection fraction, and J0923 and J1200 demonstrate that a V-shaped continuum can coexist with a radio detection.

%===================================================================
\section{Discussion}
\label{sec:discuss}
%===================================================================

% \paragraph{Excluded and allowed configurations.} 
The detections of J0923 and J1200 provide direct observational tests of the free--free attenuation predicted for a compact ionised column \citep{2026aask.bookE..77M,2026arXiv260920913T}. 
If these objects possess the compact, cool scattering layers adopted here, placing the detected radio components behind their full electron columns is incompatible with the 144\,MHz detections for any plausible intrinsic radio power (Sect.~\ref{sec:joint}). 
The components then lie outside an otherwise spherical layer or their emission reaches the observer along lower-column paths, and both configurations remain allowed. For J1047, absorption by gas with an emission measure several orders of magnitude below that of a compact $10^{24}\,{\rm cm^{-2}}$ column fits the turnover, and the spectrum leaves the thickness and distance of the absorber unconstrained. The radio data thus constrain the electron columns along the sight-line of the detected components.

% \paragraph{Strengthening the test.} 
A physical fit to the electron-scattering wings of a single source, which accounts for the degeneracy among electron temperature, scattering optical depth, and kinematic broadening, would replace the adopted column in Eq.~(\ref{eq:em_opt}) with an inferred range. Broadband radio spectra could constrain $\tau_{\max}$ and provide an observational upper bound on EM. If ${\rm EM}_{\rm min,opt}$ still exceeded ${\rm EM}_{\rm max,radio}$, the same-column configuration would be excluded for that object independently of model-adopted columns. High-resolution radio observations would establish whether the detected emission originates from the nuclear region.

% \paragraph{A changing radio state.}\label{sec:evolution} 
J0923 brightened on a decade timescale \citep{An2026a}, so the radio state of an individual analogue can evolve. The sample spans three states, namely no radio component above the survey limits, a spectrum suppressed below a few GHz (J1047), and radio detection alongside a V-shaped continuum (J0923, J1200). These may reflect evolution of the radio source or the ionised gas, or different sightlines through a stable geometry. A thinning FFA screen lowers the turnover frequency without source growth, whereas SSA in an expanding source lowers the peak frequency as the size grows. Multi-epoch broadband spectra and VLBI observations can separate FFA from SSA in J1047 (Sect.~\ref{sec:j1047}).

% \paragraph{Implications for high-redshift LRDs.} 
At $z\simeq6$, an observed frequency of 1.4--3\,GHz corresponds to 10--21\,GHz in the rest frame. For the same EM, $T_e$, and screen geometry, the homogeneous FFA fit to J1047 gives $\tff\simeq4.6$ at rest-frame 1\,GHz, falling to $\tff\simeq0.04$--$0.008$ at 10--21\,GHz, so a J1047-like screen is nearly transparent to the centimetre-wave surveys of high-redshift LRDs. The dense scattering cocoon stays opaque, since transparency at 10--21\,GHz for $T_e=10^4$\,K and $\Delta R=0.3$\,pc requires $\Ne\lesssim(3$--$7)\times10^{22}\,{\rm cm^{-2}}$ (Eq.~\ref{eq:Ne_limit}), far below $\Ne\sim10^{24}\,{\rm cm^{-2}}$. The stacked $3\sigma$ limits for 919 LRDs, $3\sigma$ mean-stack bounds of 8.8 $\mu$Jy in VLASS and 13.3 $\mu$Jy in FIRST (median-stack bounds 10.8 and 17.7 $\mu$Jy) \citep{2025A&A...693L...2P}, and the GOODS-N stack limit $L_{5\,{\rm GHz}}<2\times10^{39}\ergs$ \citep{2026A&A...706A.372M} constrain the average radio emission from components outside the dense cocoon or visible through low-column paths. A source hidden behind the full column can be indistinguishable from an intrinsically weak one. The local detections show that an LRD-like optical spectrum can coexist with a radio component that avoids the full compact column, if such a column is present. The present opacity bound applies to a component behind the full compact, cool column. A population-mean prediction additionally requires the incidence of visible components, spectral uncertainty, selection, and redshift distribution.

%===================================================================
\section{Summary}
\label{sec:summary}
%===================================================================

\noindent
We tested whether radio components detected in local LRD analogues can lie behind the full compact, cool electron columns invoked to explain the scattering wings of LRD broad lines. For fixed column and path length, a uniform screen at the maximum electron temperature gives the minimum free--free opacity, so the bound holds for any density structure. For $T_e=10^4$\,K and $\Delta R=0.3$\,pc, the $\tff=1$ column at rest-frame 3\,GHz is $\simeq10^{22}\,{\rm cm^{-2}}$, two orders of magnitude below $10^{24}\,{\rm cm^{-2}}$.

The 144\,MHz detections of the V-shaped analogues J0923 and J1200 allow an emission measure more than two orders of magnitude below that of a $10^{24}\,{\rm cm^{-2}}$ column for $T_e\leq10^6$\,K and $\Delta R\leq1$\,pc, even for an intrinsic radio power $\sim2\times10^4$ times the observed one. If these objects contain such columns, their detected components lie outside the dense layer or are viewed along much lower-column paths. The low-frequency suppression of J1047 is compatible with free--free absorption by gas of much lower emission measure or with synchrotron self-absorption, and V-shaped and non-V-shaped BLGPs show no significant difference in radio detection fraction. Source-specific fits of the scattering wings combined with broadband radio spectra can test the shared-column configuration on an object-by-object basis.

\begin{acknowledgements}
AW is supported by the National Natural Science Foundation of China  (Grant Number of 12503018). This work uses archival data from LoTSS, VLASS, ASKAP, FIRST, NVSS, the Chandra Data Archive and Source Catalog (CXC), the 4XMM-DR14 catalogue (XMM-Newton SSC), eRASS1 (eROSITA-DE Consortium), VizieR (CDS), and HEASARC (NASA/GSFC).
\end{acknowledgements}

\bibliographystyle{aa} \bibliography{ref}

%===================================================================
% APPENDIX
%===================================================================
\begin{appendix}

\section{Free--free opacity constraint}
\label{app:freefree}

In the Rayleigh--Jeans regime \citep{1967ApJ...147..471M}, the free--free optical depth for hydrogen-dominated ionised gas is
\begin{equation}
\tff \simeq 3.28\times10^{-7}
\left(\frac{T_e}{10^4\,{\rm K}}\right)^{-1.35}
\left(\frac{\nu}{\rm GHz}\right)^{-2.1}
{\rm EM}\,,
\label{eq:tff}
\end{equation}
where ${\rm EM}=\ne^2\Delta R$ in ${\rm pc\,cm^{-6}}$. For an electron column $\Ne=\ne\Delta R$, ${\rm EM}=\Ne^2/(\Delta R_{\rm pc}\times9.52\times10^{36})$, and the optical depth becomes
\begin{equation}
\tff \simeq 3.45\times10^{4}
\left(\frac{\Ne}{10^{24}\,{\rm cm^{-2}}}\right)^{2}
\left(\frac{\Delta R}{1\,{\rm pc}}\right)^{-1}
\left(\frac{T_e}{10^4\,{\rm K}}\right)^{-1.35}
\left(\frac{\nu}{\rm GHz}\right)^{-2.1},
\label{eq:tff_Ne}
\end{equation}
where $1\,{\rm pc}^2=9.52\times10^{36}\,{\rm cm^2}$ (with $\Ne$ in ${\rm cm^{-2}}$ and $\Delta R$ in pc, the factor converts $\ne=\Ne/\Delta R$ to ${\rm cm^{-3}}$). Setting $\tff=1$ gives the critical column in Eq.~(\ref{eq:Ne_limit}). Below we examine how density structure, path length, and temperature affect this constraint.

\subsection{The uniform screen gives the minimum opacity}
\label{app:freefree_cs}

For an arbitrary electron-density distribution $\ne(l)$ along a path of length $\Delta R$,
\begin{equation}
\Ne=\int_0^{\Delta R}\ne\,dl,
\qquad
{\rm EM}=\int_0^{\Delta R}\ne^2\,dl .
\end{equation}
The Cauchy--Schwarz inequality gives
\begin{equation}
{\rm EM}\geq\frac{\Ne^2}{\Delta R},
\label{eq:cs}
\end{equation}
with equality only for uniform density. The homogeneous screen used in Sect.~\ref{sec:opacity} therefore gives the minimum emission measure and minimum $\tff$ for a specified $\Ne$ and $\Delta R$. If the electrons occupy a fraction $f_V$ of the path, ${\rm EM}=\Ne^2/(f_V\Delta R)$, increasing the free--free opacity by $1/f_V$. A temperature gradient does not remove this lower bound. Since $\tff\propto\int \ne^2 T_e^{-1.35}\,dl$, replacing $T_e(l)$ by the maximum temperature along the path gives a lower limit to the opacity. Clumping or temperature structure can therefore make a sightline carrying a fixed column more opaque, but cannot make it less opaque than a uniform screen evaluated at the maximum $T_e$. A porous medium becomes radio-transparent only along rays whose electron columns are themselves reduced.

\subsection{Path length and temperature}
\label{app:freefree_T}

For a column $\Ne\sim10^{24}\,{\rm cm^{-2}}$, transparency at frequency $\nu$ requires
\begin{equation}
\Delta R_{\rm pc}
\left(\frac{T_e}{10^4\,{\rm K}}\right)^{1.35}
\gtrsim
3.4\times10^{3}
\left(\frac{\Ne}{10^{24}\,{\rm cm^{-2}}}\right)^{2}
\left(\frac{\nu}{3\,{\rm GHz}}\right)^{-2.1}.
\label{eq:transparency}
\end{equation}
At $T_e=10^4$\,K, a $10^{24}\,{\rm cm^{-2}}$ column would require $\Delta R\gtrsim3$\,kpc to become transparent at 3\,GHz, far larger than a light-day-scale scattering region. We adopt $\Delta R\lesssim0.1$\,pc as a conservative scale for the compact cocoon (Fig.~\ref{fig:opacity}).

Temperature matters most, because $\tff\propto T_e^{-1.35}$. The optical photosphere and the Balmer and He\,{\sc i} absorption favour cool gas, and we treat $T_e=10^6$\,K as a conservative case and $10^7$\,K as an extreme one. At $10^6$\,K the opacity drops by $\simeq500$, but $\Delta R\leq0.1$\,pc still gives $\tff(3\,{\rm GHz})\gtrsim70$ (Table~\ref{tab:robust}). At $10^7$\,K, transparency at 3\,GHz requires $\Delta R\gtrsim0.3$\,pc, well outside the photospheric scale. At the rest-frame 0.19\,GHz of the J0923 and J1200 detections, the exact Gaunt factor lowers $\tff$ by 24\% and 37\% at $10^6$ and $10^7$\,K, which reduces $Q$ by the same fractions (from 225 to about 170 for $\Ne=10^{24}\,{\rm cm^{-2}}$, $\Delta R=1$\,pc, $T_e=10^6$\,K).

The critical column (Eq.~\ref{eq:Ne_limit}) depends more weakly on these quantities. At 3\,GHz it ranges from $\sim10^{21}\,{\rm cm^{-2}}$ ($\Delta R=3\times10^{-3}$\,pc, $T_e=10^4$\,K) to $\sim4\times10^{23}\,{\rm cm^{-2}}$ ($\Delta R=1$\,pc, $T_e=10^6$\,K). An observed 144\,MHz detection of a source at redshift $z$ lowers the limit by $[0.144(1+z)/3]^{1.05}$ relative to rest-frame 3\,GHz, or $\simeq0.055$ (a factor of about 18) for J0923 and J1200.

\subsection{What a radio detection constrains}
\label{app:freefree_paths}

\begin{figure}
\centering
\includegraphics[width=\columnwidth]{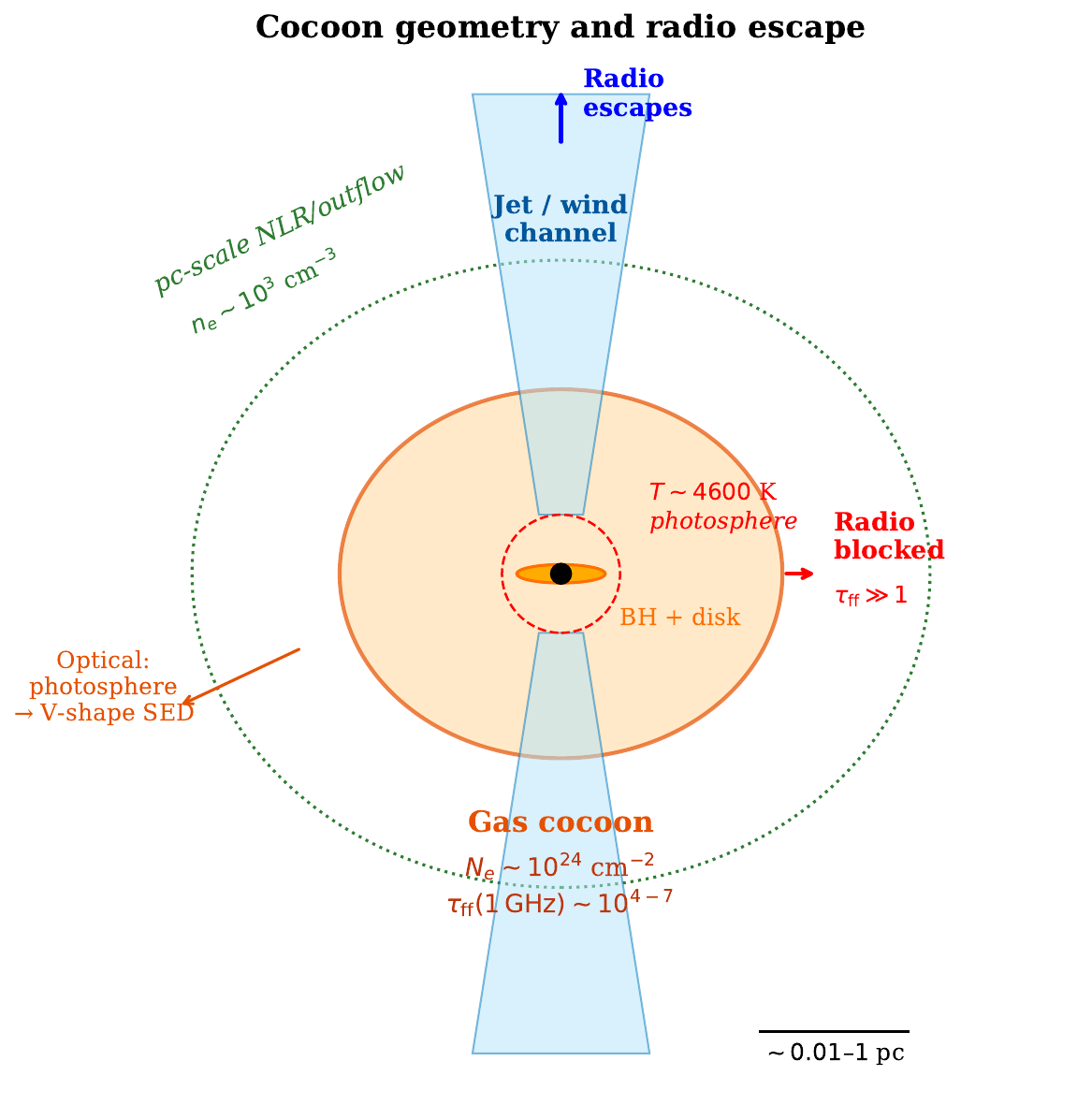}
\caption{Schematic of the cocoon geometry. The dense scattering layer ($\Ne\sim10^{24}\,{\rm cm^{-2}}$, orange) absorbs radio photons that traverse its full column. The optical lines and continuum form in the ionised gas, and the line photons are broadened by electron scattering. For $T_e=10^4$\,K and $\Delta R\lesssim0.3$\,pc, a path transparent at rest-frame 3\,GHz has $\Ne\lesssim10^{22}\,{\rm cm^{-2}}$. The emitter (a jet, wind shock, or starburst) either lies outside the cocoon or escapes through a low-column channel. Pc-scale ionised gas (green dotted outline) can impose additional free--free absorption on the escaping radio, as possibly in J1047.}
\label{fig:schematic}
\end{figure}

The free--free argument constrains the column traversed by the radio photons rather than the location of the emitter. The emitter may lie outside the dense scattering layer, or further in but viewed through a low-column channel such as a polar opening or a porous path through an anisotropic cocoon (Fig.~\ref{fig:schematic}). A spherically symmetric cocoon with a radio component outside it yields a detection as readily as a porous or non-spherical one. For partial covering, or for emission mixed with the absorbing gas, the escaping photons originate in a low-column surface layer and do not bound the total column. VLBI size and brightness temperature, spectral turnovers, and variability locate the radio-emitting region. Electron scattering ($\tau_{\rm es}\simeq0.67\,\Ne/10^{24}\,{\rm cm^{-2}}$) redistributes rather than absorbs radio photons, so the radio constraint measures the electron column along the radio path. Radio observations therefore show which radio-emitting components have access to transparent paths.

\subsection{Radio luminosity relative to a standard corona}
\label{app:corona}

For J0923 and J1200, the $L_X$ expected for a typical AGN of the same $\Lbol$ (not a measured intrinsic X-ray luminosity) is $L_X\simeq\Lbol/20\sim10^{43}\ergs$ (Appendix~\ref{app:xray}). With $L_R/L_X\sim10^{-5}$ \citep{2008MNRAS.390..847L}, this gives $L_\nu(5\,{\rm GHz})\approx2\times10^{28}\ergshz$, a factor of 17--53 (for the adopted $k_{\rm bol}$, $\alpha=-0.7$, and Table~\ref{tab:master} values of $\Rbol$) below the observed luminosities. Matching the radio output with the same ratio would require an intrinsic coronal $L_X\sim(2$--$4)\times10^{44}\ergs$, comparable to or larger than the bolometric luminosity $\Lbol$. Any nuclear radio emission from a corona, jet, or wind shock would be strongly absorbed if viewed through the full scattering column.

\begin{table}[!htbp]
\caption{Free--free optical depth at 3\,GHz for a column of $\Ne=10^{24}\,{\rm cm^{-2}}$. Parenthetical values give the frequency, in GHz, at which $\tff=1$.\label{tab:robust}}
\centering
\footnotesize
\setlength{\tabcolsep}{2.5pt}
\begin{tabular}{lcccc}
\hline\hline
$\Delta R$ (pc) & $10^4$\,K & $10^5$\,K & $10^6$\,K & $10^7$\,K \\
\hline
$3\times10^{-3}$ & $1.1\times10^{6}$ (2300) & $5.1\times10^{4}$ (520) & $2.3\times10^{3}$ (120) & 100 (27) \\
$10^{-2}$ & $3.4\times10^{5}$ (1300) & $1.5\times10^{4}$ (300) & 680 (67) & 31 (15) \\
0.1 & $3.4\times10^{4}$ (430) & $1.5\times10^{3}$ (99) & 68 (22) & 3.1 (5) \\
1 & $3.4\times10^{3}$ (145) & 150 (33) & 6.8 (7) & 0.3 (2) \\
\hline
\end{tabular}
\tablefoot{Uniform slab with the $T_e^{-1.35}$ scaling of Eq.~(\ref{eq:tff}). The corresponding densities are $\ne=\Ne/\Delta R\simeq1.1\times10^{8}$, $3.2\times10^{7}$, $3.2\times10^{6}$, and $3.2\times10^{5}\,{\rm cm^{-3}}$. The exact Gaunt factor lowers the $10^6$ and $10^7$\,K entries by $\lesssim15\%$ and $\lesssim30\%$ at 3\,GHz. All entries are lower limits for clumped media (Eq.~\ref{eq:cs}). The cocoon parameter space of Fig.~\ref{fig:opacity} corresponds to $\Delta R\lesssim0.1$\,pc.}
\end{table}

\section{Sample construction and multi-wavelength data}
\label{app:sample}

The 44 local LRD analogues used in this work are assembled from six published samples selected through different combinations of broad Balmer emission, compact morphology, V-shaped UV-to-optical continua, large Balmer decrements, absorption features, and X-ray weakness. The catalogue is a union of literature selections. We use it to characterise the range of radio properties and to compare matched subsamples. After removing overlaps, the combined sample contains 44 unique sources. Table~\ref{tab:master} lists the quantities used in the analysis. The full machine-readable catalogue, including the individual radio measurements and uncertainties, is shared with Paper~II and includes coverage, threshold, and measurement-provenance fields  \citep[see][]{Wang2026}.

\subsection{Parent catalogues}

\paragraph{Lin+26a (3 sources).} \citet{2026ApJ...997..364L} reported three local LRD analogues at $z=0.10$--$0.22$ selected from Sloan Digital Sky Survey (SDSS) spectroscopy and followed up with LBT/MODS and Magellan/FIRE. All three show V-shaped SEDs, broad Balmer emission, and blue-shifted He\,{\sc i} absorption, with Balmer absorption in two. The two with X-ray observations, J1025 and J1047, are X-ray weak (Appendix~\ref{app:xray}). $L_{5100}$ is measured from the MODS spectra, and $M_{\rm BH}$ is estimated from broad H$\alpha$.

\paragraph{Lin+25 (7 V-shaped BLGPs).} \citet{2024SCPMA..6709811L} identified 59 Green Peas with broad H$\alpha$ components. Applying ${\rm FWHM}_{\rm H\alpha,broad}>1000$\,km\,s$^{-1}$, \citet{2025ApJ...980L..34L} retained 19 broad-line Green Pea galaxies (BLGPs) and identified seven of them as V-shaped local LRD analogues, selected by $\beta_{\rm UV}<-0.37$ and $\beta_{\rm opt}\gtrsim0$. Published $M_{\rm BH}$ and $M_*$ values are adopted from their Table~1. The remaining twelve BLGPs form the non-V-shaped comparison sample used in Sect.~\ref{sec:control}. Two of them, J1405+5155 and J1605+4405, lack GALEX coverage and therefore have no measured $\beta_{\rm UV}$. J1405 has $\beta_{\rm opt}<0$ and is not V-shaped regardless of its UV slope. J1605 has a red optical slope ($\beta_{\rm opt}=1.93$), so its classification is untested. Removing it from the comparison gives $7/10$ versus $3/7$ at 144\,MHz and $4/11$ versus $2/7$ at 3\,GHz for ${\rm S/N}\geq3$.

\paragraph{Lin+26b (27 DESI DR1 LRDs).} \citet{2026arXiv260521574L} identified 27 LRD analogues at $z=0.2$--$0.9$ from Dark Energy Spectroscopic Instrument (DESI) DR1 by spectroscopic template matching to high-redshift LRD spectra. Rest-frame $L_{5100}$ is measured from the DESI spectra.

\paragraph{Ding+26, Chen+26, and Park+26.} \citet{2026A&A...709L..11D} identified five DESI LRD analogues at $z=0.25$--$0.38$. We adopt their black hole masses and derive $L_{5100}$ from their photometry (Appendix~\ref{app:lbol}). For J2048 ($z=0.433$, \citealt{2026ApJ...999...30C}), with broad H$\alpha$ ${\rm FWHM}=9520\pm350$\,km\,s$^{-1}$, we adopt the extinction-corrected $\Lbol=10^{13.6\pm0.4}\,L_\odot$ ($A_V=6.28\pm0.27$). Seven of the eight DESI DR1 LRDs of \citet{2026arXiv260514233P} overlap with Lin+26b, and the remaining source, J0716+5433, has no $L_{5100}$ measurement and therefore no $\Lbol$ estimate in this work.

\subsection{Uniform bolometric luminosity}
\label{app:lbol}

For sources with a measured rest-frame 5100\,\AA\ monochromatic luminosity, we place the sample on a common bolometric scale using
\begin{equation}
\Lbol = 9.8\,L_{5100},
\label{eq:lbol}
\end{equation}
following \citet{2004MNRAS.352.1390M}. In LRDs, the 5100\,\AA\ continuum may include substantial photospheric emission \citep{2025ApJ...988L..22I}, so we use Eq.~(\ref{eq:lbol}) as a uniform luminosity scale for comparing radio-to-bolometric ratios across the sample. For the seven V-shaped BLGPs without $L_{5100}$, $\Lbol$ is estimated from broad H$\alpha$ luminosity using the \citet{2005ApJ...630..122G} relation. For the five \citet{2026A&A...709L..11D} sources, $L_{5100}$ is interpolated from their Galactic-extinction-corrected DESI Legacy Survey $r$- and $i$-band photometry. The published broad H$\alpha$ luminosity of J0929 exceeds the integrated luminosity of the $z$ band that contains H$\alpha$, so an H$\alpha$-based $\Lbol$ is not used for this subsample.

\subsection{Radio efficiency}
\label{app:rbol}

The conventional radio-loudness parameter $\mathcal{R}=f_{5\,{\rm GHz}}/f_{4400\,\text{\AA}}$ \citep{1989AJ.....98.1195K} is poorly matched to the present sample, because 5\,GHz measurements are not uniformly available and the optical continuum can be dominated by the reprocessing photosphere rather than by the intrinsic accretion-disc continuum \citep{2026Natur.649..574R,2026NatAs..10.1261I}. We therefore define
\begin{equation}
\Rbol \equiv \frac{\nu L_\nu(1.4\,{\rm GHz})}{\Lbol}.
\end{equation}
Because the 5100\,\AA\ continuum of LRDs may be dominated by reprocessed emission, $\Rbol$ serves as comparative radio-to-bolometric indicator rather than a direct measure of intrinsic radio loudness.

Radio measurements use the same signed forced-photometry release as Paper~II \citep{Wang2026}. LoTSS DR2/DR3 provides 35 usable 144 MHz measurements; missing coverage is not a non-detection. At the fixed optical position, a restoring-beam Gaussian is fitted with a local background. Statistical errors are the largest of the local rms, blank-position fit scatter, and fitting-variant scatter. Calibration systematics are treated separately. VLASS epochs 1–-3 are fitted at native resolution and their amplitudes combined by inverse statistical variance, rather than measuring an image-plane co-addition. ASKAP, FIRST, and NVSS provide ancillary constraints.
% They comprise 144\,MHz LoTSS DR2, ASKAP measurements at 0.9--1.7\,GHz from RACS \citep[Rapid ASKAP Continuum Survey;][]{2020PASA...37...48M,2021PASA...38...58H} and VAST \citep{2021PASA...38...54M}, 1.4\,GHz FIRST \citep{1995ApJ...450..559B} and NVSS \citep{1998AJ....115.1693C}, and 3\,GHz VLASS, all measured at the optical positions. For radio detections, $L_\nu(1.4\,{\rm GHz})$ is derived from the nearest detected band using $S_\nu\propto\nu^{-0.7}$ with a K-correction, and J1047 uses its published 6\,GHz core flux density of 0.117\,mJy \citep{2026A&A...707L..17R}. 
For $R_{\rm bol}$, we use the nearest detected primary census band (LoTSS or VLASS) with $\alpha = -0.7$ and a K-correction, except J1047, which uses its published 6 GHz core (0.117 mJy; \citealt{2026A&A...707L..17R}). J1200 uses its nominal $3\sigma$ VLASS measurement.
Non-detections are reported as $3\sigma$ upper limits, and their signed amplitudes, including negative values, are retained only for stacking (Appendix~\ref{app:stacking}). Luminosities and sizes adopt a flat $\Lambda$CDM cosmology with $H_0=70\,{\rm km\,s^{-1}\,Mpc^{-1}}$ and $\Omega_m=0.3$.

\subsection{Derived black hole masses}
\label{app:derived}

Where a black hole mass has been published, we adopt that value. For eight sources with broad H$\alpha$ luminosity and line width but no published $M_{\rm BH}$, we use the single-epoch estimator of \citet{2015ApJ...813...82R},
\begin{equation}
\begin{split}
\log\!\left(\frac{M_{\rm BH}}{M_\odot}\right) = 6.57 + 0.47\log\!\left(\frac{L_{\rm H\alpha,broad}}{10^{42}\,{\rm erg\,s^{-1}}}\right) \\ {}+ 2.06\log\!\left(\frac{{\rm FWHM}_{\rm H\alpha}}{10^3\,{\rm km\,s^{-1}}}\right).
\end{split}
\label{eq:mbh_rv15}
\end{equation}
The intrinsic scatter of the virial calibration is approximately 0.4\,dex. These virial estimates assume Doppler broadening and overestimate $M_{\rm BH}$ if electron scattering dominates the line width. Masses derived in this work are flagged ``RV15'' in Table~\ref{tab:master}.

\subsection{Stacking of non-detections}
\label{app:stacking}

For each frequency, we compute the inverse-variance weighted mean of the signed forced-photometry measurements for sources individually undetected in that band,
\begin{equation}
\bar{S} = \frac{\sum_i w_i S_i}{\sum_i w_i}, \qquad w_i=\sigma_i^{-2},
\label{eq:stack}
\end{equation}
with uncertainty $(\sum_i w_i)^{-1/2}$. Retaining negative amplitudes preserves the signed noise fluctuations and avoids the positive bias introduced by replacing non-detections with upper limits.

At 144\,MHz, 30 sources have usable LoTSS measurements and no individual LoTSS detection. Their weighted mean is $\bar{S}_{144}=0.008\pm0.015$\,mJy ($0.5\sigma$). At 3\,GHz, the 38 sources outside the nominal $3\sigma$ VLASS group give $\bar{S}_{3}=0.016\pm0.013$\,mJy ($1.2\sigma$). Neither stack yields a significant mean signal. We note that these are summaries of the selected non-detections, not estimates of the full-sample mean or a typical radio state.

For Fig.~\ref{fig:Rbol}, the LoTSS stack limit, $\bar{S}_{144}+3\sigma=0.053$\,mJy, is converted to a representative radio-efficiency limit using the median properties of the 30 stacked sources, $z_{\rm med}=0.419$ and $\log\Lbol\simeq44.24$, with $\alpha=-0.7$, giving $\log\Rbol<-6.31$. This quantity represents the stack at the characteristic source properties and is not an average of the individual $\Rbol$ values.

\subsection{X-ray data}
\label{app:xray}

X-ray information is auxiliary and enters none of the radio analysis. We cross-matched the sample (10$\arcsec$ radius) with the Chandra Source Catalog~2.1 \citep{2024ApJS..274...22E}, 4XMM-DR14 \citep{2020A&A...641A.136W}, and the first eROSITA all-sky survey (eRASS1, \citealt{2024A&A...682A..34M}), and used the HEASARC pointing archives to establish which sources were actually observed. Table~\ref{tab:master} records the outcome in three states. A detection ($\checkmark$) or a non-detection ($\times$) is assigned only where the source was covered by one of these data sets or by a published measurement. The remaining 24 sources have no valid X-ray information (``\ldots''), which is neither a detection nor a non-detection. 
% Pointings that postdate these catalogues or have not yet been executed are not used.

\paragraph{Detection.} J1047 has a catalogued Chandra counterpart (2CXO J104755.9+073950), with $5.6^{+2.8}_{-2.7}$ net counts in 53\,ks and $L_{2\text{--}10\,{\rm keV}}\simeq8\times10^{40}\ergs$ \citep{2026ApJ...997..364L}. The detection is significant at only $\lesssim2\sigma$, so we mark it as tentative, and its luminosity is $\sim2$\,dex below the expected value (see below).

\paragraph{Non-detections.} J1025 was not detected by Chandra in 4.9\,ks ($L_{2\text{--}10\,{\rm keV}}<1.1\times10^{41}\ergs$), implying an $L_X/L_{{\rm H}\alpha}$ ratio at least 1--2\,dex below that of type-1 AGN \citep{2026ApJ...997..364L}. J1222 was not detected by Chandra in 11\,ks ($L_{0.5\text{--}7\,{\rm keV}}<1.5\times10^{42}\ergs$ at $2\sigma$, \citealt{2024SCPMA..6709811L}). Of the 19 sources inside the eRASS1 footprint, J1047 and J1025 are discussed above and J1200 is treated separately. For five of the remaining 16 (J0852, J0928, J0929, J1022, J1051) the 0.5--2\,keV limit lies at or below the expected $L_X$, and these are marked $\times$. For the other 11 the limit lies above the expected $L_X$ and is not constraining. These are marked $\times^{s}$ and say nothing about X-ray weakness. The eRASS1 limits adopt the survey-average 3$\sigma$ flux limit of $5\times10^{-14}\,{\rm erg\,cm^{-2}\,s^{-1}}$ (0.2--2.3\,keV), not source-specific sensitivities.

\paragraph{Expected $L_X$.} 
% We take $L_{2\text{--}10\,{\rm keV}}=\Lbol/k_{\rm bol}$ with $k_{\rm bol}=20$, 50, and 100 for $\log\Lbol<44.5$, $<45.5$, and larger, respectively, representative of the bolometric corrections of \citet{2004MNRAS.351..169M} and \citet{2020A&A...636A..73D}, and assume the same value in the 0.5--2\,keV band. 
We take $L_{2\text{--}10\,{\rm keV}}=\Lbol/k_{\rm bol}$ with $k_{\rm bol}=20$, 50, and 100 for $\log\Lbol<44.5$, $44.5\leq\log\Lbol<45.5$, and $\log\Lbol\geq45.5$, respectively, representative of the bolometric corrections of \citet{2004MNRAS.351..169M} and \citet{2020A&A...636A..73D}, and approximate $L_{0.5\text{--}2\,{\rm keV}}\simeq L_{2\text{--}10\,{\rm keV}}$. 
This is the luminosity of a typical unobscured AGN of the same $\Lbol$, not a measurement of the intrinsic X-ray power of an LRD. The classification against the survey-average eRASS1 limit is approximate for the five boundary sources.

\paragraph{Sources without a classification.} J1200 has a single eRASS1 counterpart (1eRASS J120051.9+331235, 4 counts, $6\arcsec$ offset) at the expected level, which we treat as a marginal counterpart (``m'') and count neither as a detection nor as a non-detection. J0923 lies outside the eRASS1 footprint and has no counterpart in the catalogues above, and its parent catalogue lists no Chandra or XMM-Newton coverage \citep{2024SCPMA..6709811L}. J0923 and the 23 other sources without valid X-ray observation therefore carry no X-ray classification, and the V-shaped BLGPs are not assigned to either the X-ray weak or the X-ray normal class.

In summary, the public data indicate X-ray weakness in J1025 and J1047, provide shallow or no constraints for most of the remaining sources, and are consistent with the dense-cocoon expectation of strong X-ray attenuation. A census of the X-ray visibility of local LRDs requires dedicated X-ray observations.

\begin{table*}
\caption{Multi-wavelength properties of 44 local LRD analogues. The radio column gives detection status in the shared LoTSS/VLASS census, including the published VLA detection of J1047. The X-ray column is auxiliary and gives a status from archival X-ray data, defined in the table notes and in Appendix~\ref{app:xray}.\label{tab:master}}
\centering
\footnotesize\renewcommand{\arraystretch}{0.92}
\begin{tabular}{lcccccccccc}
\hline\hline
Source & $z$ & Subsample & $\log \Lbol$ & $\log M_{\rm BH}$ & FWHM$_{\rm H\alpha}$ & H$\alpha$/H$\beta$ & $\log \Rbol$ & Radio & X-ray & $M_{\rm BH}$ \\
 & & & (erg\,s$^{-1}$) & ($M_\odot$) & (km\,s$^{-1}$) & (broad) & & det. & status & ref. \\
\hline
J0009+0811 & 0.372 & L26b & 44.11 & \ldots & \ldots & \ldots & $<-5.56$ & $\times$ & \ldots & \ldots \\
J0035+0848 & 0.242 & L25 & 44.07 & 6.99 & \ldots & \ldots & $<-5.91$ & $\times$ & \ldots & L25 \\
J0129+0628 & 0.247 & L26b & 43.89 & 6.40 & 748 & 11.0 & $<-5.73$ & $\times$ & \ldots & RV15 \\
J0243+0349 & 0.458 & L26b & 44.03 & \ldots & \ldots & \ldots & $<-5.21$ & $\times$ & \ldots & \ldots \\
J0716+5433 & 0.291 & P26 & \ldots & 6.97 & 1258 & 7.7 & \ldots & $\times$ & \ldots & RV15 \\
J0826$-$0100 & 0.627 & L26b & 44.85 & \ldots & \ldots & \ldots & $<-5.24$ & $\times$ & $\times^{s}$ & \ldots \\
J0829+1312 & 0.399 & L26b & 44.58 & 6.85 & 944 & 7.4 & $<-5.86$ & $\times$ & $\times^{s}$ & RV15 \\
J0850+3030 & 0.280 & L25 & 43.41 & 6.29 & \ldots & \ldots & $<-5.18$ & $\times$ & $\times^{s}$ & L25 \\
J0852$-$0411 & 0.370 & D26 & 45.00 & 7.97 & \ldots & \ldots & $<-5.53$ & $\times$ & $\times$ & D26 \\
J0923+4157 & 0.323 & L25 & 44.22 & 6.86 & \ldots & \ldots & $-4.70$ & $\checkmark$ & \ldots & L25 \\
J0928+2921 & 0.293 & L25 & 44.69 & 7.76 & \ldots & \ldots & $-5.99$ & $\checkmark$ & $\times$ & L25 \\
J0929$-$0251 & 0.340 & D26 & 45.07 & 8.96 & \ldots & \ldots & $<-5.68$ & $\times$ & $\times$ & D26 \\
J0944$-$0249 & 0.662 & L26b & 45.21 & \ldots & \ldots & \ldots & $<-5.13$ & $\times$ & $\times^{s}$ & \ldots \\
J1017+3114 & 0.671 & L26b & 44.65 & \ldots & \ldots & \ldots & $<-5.61$ & $\times$ & $\times^{s}$ & \ldots \\
J1022+0841 & 0.223 & L26a & 44.24 & 6.74 & 805 & 33.5 & $<-6.27$ & $\times$ & $\times$ & L26a \\
J1025+1402 & 0.101 & L26a & 43.87 & 6.52 & 956 & 16.9 & $<-6.55$ & $\times$ & $\times$ & L26a \\
J1025+5028 & 0.882 & L26b & 45.30 & \ldots & \ldots & \ldots & $<-6.13$ & $\times$ & \ldots & \ldots \\
J1042+3721 & 0.608 & L26b & 44.91 & \ldots & \ldots & \ldots & $<-6.08$ & $\times$ & $\times^{s}$ & \ldots \\
J1047+0739 & 0.168 & L26a & 44.26 & 6.76 & 952 & 8.3 & $-5.73$ & $\checkmark$ & $\checkmark^{t}$ & L26a \\
J1051$-$0626 & 0.380 & D26 & 45.24 & 8.02 & \ldots & \ldots & $-5.31$ & $\checkmark$ & $\times$ & D26 \\
J1056+2754 & 0.462 & L26b & 44.29 & \ldots & \ldots & \ldots & $<-5.60$ & $\times$ & $\times^{s}$ & \ldots \\
J1059+3149 & 0.499 & L26b & 44.58 & \ldots & \ldots & \ldots & $<-5.89$ & $\times$ & $\times^{s}$ & \ldots \\
J1119+0219 & 0.468 & L26b & 44.20 & \ldots & \ldots & \ldots & $<-5.11$ & $\times$ & $\times^{s}$ & \ldots \\
J1137+5520 & 0.436 & L26b & 43.99 & 6.56 & 891 & 8.9 & $<-5.58$ & $\times$ & \ldots & RV15 \\
J1154+0650 & 0.269 & L25 & 44.52 & 7.62 & \ldots & \ldots & $<-5.50$ & $\times$ & $\times^{s}$ & L25 \\
J1200+3312 & 0.303 & L25 & 44.38 & 8.41 & \ldots & \ldots & $-5.23$ & $\checkmark$ & m & L25 \\
J1222+3602 & 0.301 & L25 & 44.09 & 6.69 & \ldots & \ldots & $<-5.89$ & $\times$ & $\times$ & L25 \\
J1321$-$0214 & 0.224 & L26b & 43.59 & \ldots & \ldots & \ldots & $<-4.50$ & $\times$ & $\times^{s}$ & \ldots \\
J1343+3934 & 0.293 & L26b & 43.89 & 6.11 & 588 & 5.3 & $<-5.72$ & $\times$ & \ldots & RV15 \\
J1423+5202 & 0.624 & L26b & 44.80 & \ldots & \ldots & \ldots & $<-5.93$ & $\times$ & \ldots & \ldots \\
J1502+0250 & 0.291 & L26b & 43.95 & 6.27 & 685 & 13.2 & $<-5.34$ & $\times$ & \ldots & RV15 \\
J1611+0917 & 0.695 & L26b & 44.93 & \ldots & \ldots & \ldots & $<-5.64$ & $\times$ & \ldots & \ldots \\
J1620+3148 & 0.410 & L26b & 43.29 & \ldots & \ldots & \ldots & $<-4.76$ & $\times$ & \ldots & \ldots \\
J1641+0708 & 0.535 & L26b & 44.62 & \ldots & \ldots & \ldots & $<-5.32$ & $\times$ & \ldots & \ldots \\
J1642+0426 & 0.729 & L26b & 45.21 & \ldots & \ldots & \ldots & $<-5.07$ & $\times$ & \ldots & \ldots \\
J1646+1426 & 0.707 & L26b & 45.17 & \ldots & \ldots & \ldots & $<-6.14$ & $\times$ & \ldots & \ldots \\
J1654+0337 & 0.641 & L26b & 45.04 & \ldots & \ldots & \ldots & $<-5.03$ & $\times$ & \ldots & \ldots \\
J1717+3807 & 0.196 & L26b & 44.42 & 7.11 & 1248 & 8.9 & $-5.56$ & $\checkmark$ & \ldots & RV15 \\
J1909+5831 & 0.427 & L26b & 44.11 & 6.30 & 748 & 9.6 & $<-5.49$ & $\times$ & \ldots & RV15 \\
J2048$-$0024 & 0.433 & C26 & 47.18$^{a}$ & 10.20 & 9520 & \ldots & $-6.60^{a}$ & $\checkmark$ & \ldots & C26 \\
J2127$-$0448 & 0.584 & L26b & 44.81 & \ldots & \ldots & \ldots & $<-4.95$ & $\times$ & \ldots & \ldots \\
J2146+1028 & 0.250 & D26 & 45.25 & 7.86 & \ldots & \ldots & $-5.87$ & $\checkmark$ & \ldots & D26 \\
J2255+1542 & 0.427 & L26b & 43.99 & \ldots & \ldots & \ldots & $<-5.35$ & $\times$ & \ldots & \ldots \\
J2357$-$1128 & 0.300 & D26 & 45.33 & 6.85 & \ldots & \ldots & $<-5.54$ & $\times$ & \ldots & D26 \\
\hline
\end{tabular}
\tablefoot{Subsample abbreviations are L26a = \citet{2026ApJ...997..364L}, L25 = \citet{2025ApJ...980L..34L}, L26b = \citet{2026arXiv260521574L}, D26 = \citet{2026A&A...709L..11D}, C26 = \citet{2026ApJ...999...30C}, and P26 = \citet{2026arXiv260514233P}. $M_{\rm BH}$ reference codes denote the original publication, and RV15 marks values derived in this work using Eq.~(\ref{eq:mbh_rv15}). $\Lbol$ and $\Rbol$ are defined in Appendices~\ref{app:lbol} and \ref{app:rbol}. Upper limits are $3\sigma$, from LoTSS where a usable measurement exists and from VLASS otherwise. $^{a}$Extinction-corrected $\Lbol=10^{13.6}\,L_\odot$ from \citet{2026ApJ...999...30C} ($A_V=6.3$), so the inferred $\Rbol$ depends on the extinction correction. In the radio column, $\checkmark$ means detected and $\times$ means not detected, and J1047 is detected by the VLA only. The X-ray column is auxiliary and enters none of the radio analysis. There, $\checkmark^{t}$ means a tentative detection ($\lesssim2\sigma$), $\times$ means covered by Chandra, XMM-Newton, or eRASS1 and not detected with a limit at or below the $L_X$ expected from $\Lbol$ ($L_X=\Lbol/k_{\rm bol}$ with $k_{\rm bol}=20$, 50, 100, following \citealt{2004MNRAS.351..169M} and \citealt{2020A&A...636A..73D}), $\times^{s}$ means not detected with a shallow limit above the expected $L_X$, m means a marginal counterpart (one eRASS1 source, 4 counts, $6\arcsec$ offset) counted neither as a detection nor as a non-detection, and \ldots{} means no usable X-ray information (outside all footprints, no public pointing, or pointing not yet executed or released), so that no classification is assigned. X-ray data come from the Chandra Source Catalog, the Chandra and XMM-Newton observation logs, and the 4XMM-DR14 catalogue (all via HEASARC), from eRASS1 \citep{2024A&A...682A..34M} with the survey-average $3\sigma$ limit of $5\times10^{-14}\,{\rm erg\,s^{-1}\,cm^{-2}}$ (0.2--2.3\,keV) in place of source-specific limits, and from the literature for J1025 and J1047 \citep{2026ApJ...997..364L}. See Appendix~\ref{app:xray}.}
\end{table*}

\section{LoTSS and VLASS photometry of the non-V-shaped BLGPs}
\label{app:control}

The comparison sample consists of the twelve BLGPs from \citet{2024SCPMA..6709811L} that were not classified as V-shaped by \citet{2025ApJ...980L..34L}. The optical positions are the spectroscopic fibre positions from SDSS DR16 \citep{2020ApJS..249....3A}, or from LAMOST DR5 and DR7 for J1243$-$0242 and J2251$-$0030, which lack SDSS spectra. The V-shaped BLGPs from Paper~II are measured in the same way at both frequencies (Table~\ref{tab:control}).

\begin{table}[!t]
\renewcommand{\thetable}{C.1}\renewcommand{\theHtable}{C.1}\addtocounter{table}{-1}
\caption{LoTSS 144\,MHz and VLASS 3\,GHz photometry of the non-V-shaped BLGPs.\label{tab:control}}
\centering
\footnotesize
\setlength{\tabcolsep}{3pt}
\begin{tabular}{lccccc}
\hline\hline
Source & $z$ & $S_{144}$ & S/N & $S_{3}$ & S/N \\
 & & (mJy) & & (mJy) & \\
\hline
J0840+4707 & 0.042 & $0.28\pm0.07$ & 4.1$^{t}$ & $0.28\pm0.08$ & 3.5$^{t}$ \\
J0938+4117 & 0.280 & $0.26\pm0.06$ & 4.2$^{t}$ & $-0.04\pm0.08$ & $-0.5$ \\
J0956+4307 & 0.276 & $0.25\pm0.06$ & 4.0$^{t}$ & $0.13\pm0.08$ & 1.6 \\
J1126+3858 & 0.337 & $1.62\pm0.06$ & 27.3 & $0.37\pm0.08$ & 4.8$^{t}$ \\
J1148+1756 & 0.079 & $0.32\pm0.09$ & 3.8$^{t}$ & $0.25\pm0.09$ & 2.9 \\
J1243$-$0242 & 0.210 & \ldots & \ldots & $0.04\pm0.09$ & 0.5 \\
J1405+5155 & 0.271 & $1.51\pm0.12$ & 12.1 & $1.06\pm0.09$ & 12.4 \\
J1426+2711 & 0.356 & $13.12\pm0.37$ & 35.5 & $1.55\pm0.09$ & 17.4 \\
J1605+4405 & 0.320 & $9.22\pm0.12$ & 79.4 & $1.09\pm0.08$ & 13.2 \\
J2228+1622 & 0.226 & $0.08\pm0.09$ & 0.8 & $0.04\pm0.09$ & 0.5 \\
J2250+0000$^{a}$ & 0.081 & $-0.12\pm0.21$ & $-0.6$ & $0.07\pm0.09$ & 0.8 \\
J2251$-$0030$^{a}$ & 0.081 & $-0.20\pm0.23$ & $-0.9$ & $0.16\pm0.09$ & 1.8 \\
\hline
\end{tabular}
\tablefoot{Secure detections have ${\rm S/N}\geq5$, and $^{t}$ marks tentative detections ($3\leq{\rm S/N}<5$). VLASS observations comprise three epochs, whose amplitudes are combined by inverse statistical variance. In the V-shaped comparison sample, J0923, J0928, and J1200 are secure LoTSS detections, J0923 is a secure VLASS detection ($0.44 \pm 0.08$ mJy), and J1200 is reported as $3.0\sigma$. $^{a}$LoTSS DR3, $9''$ beam.}
\end{table}

\textit{LoTSS.} We use the forced-photometry procedure of Paper~II \citep{Wang2026}. A beam-shaped Gaussian is fitted at the fixed optical position together with a local planar background, and the flux density is the amplitude of this fit. The adopted uncertainty is the largest of the local rms, the scatter of identical fits at nearby blank positions, and the scatter among fitting windows. A second fit with the centroid left free is used only to test the positional association. A source is securely detected when ${\rm S/N}\geq5$ and the free-fit centroid lies within half a restoring beam of the optical position, and tentative when $3\leq{\rm S/N}<5$ with the same positional criterion. The LoTSS image of J1243$-$0242 contains no valid pixels at its position. J2250+0000 and J2251$-$0030 lie in the LoTSS DR3 region ($9''$ beam instead of $6''$). J1426+2711 is retained because its 13.1\,mJy component is centred $0.5''$ from the optical position and a separate 259\,mJy source lies $42''$ away, outside the fitting region. The four secure detections have free-fit centroids within $0.8''$ of the optical positions.

\textit{VLASS.} 
We use the same estimator as Paper II: a fixed-position restoring-beam Gaussian and local background are fitted to each native Quick Look image, with the same statistical error prescription, and epochs 1–3 are combined by inverse statistical variance. Image-plane averages are used for visualisation only. A secure or tentative detection requires the corresponding $5\sigma$ or $3\sigma$ threshold and a positional counterpart.
% Quick Look images from all available epochs (1--3) are regridded onto a common $1''$-pixel grid centred on the optical position and averaged, following the Paper~II procedure. The per-epoch noise levels are comparable (0.11--0.21\,mJy\,beam$^{-1}$), so equal weighting is close to inverse-variance weighting. The flux density is the value of the averaged map at the optical position, approximately equal to the integrated flux density for a source unresolved by the $\sim2.5''$ beam, and the uncertainty is the robust rms in a $30''$--$80''$ annulus. The secure and tentative S/N thresholds are as for LoTSS.

Three of the four secure LoTSS sources in the non-V-shaped sample are secure VLASS detections. J1126+3858 has a native-fit VLASS mean of $0.367 \pm 0.076$ mJy ($4.83\sigma$) and is tentative. Counting secure and tentative detections, the fractions are $8/11$ (LoTSS) and $5/12$ (VLASS) for the non-V-shaped BLGPs, compared with $3/7$ and $2/7$ for the V-shaped BLGPs under the nominal $3\sigma$ convention. The secure-only fractions are $4/11$ versus $3/7$ and $4/12$ versus $1/7$. 
% Two-sided Fisher exact tests give $p=1.0$, 0.33, 0.60, and 0.66 for the four comparisons (144\,MHz and 3\,GHz, secure and secure plus tentative).
Two-sided Fisher exact tests give $p=1.0$ and 0.33 at 144\,MHz, and 1.00 and 0.66 at 3\,GHz, for secure-only and secure-plus-tentative detections, respectively.
At 144\,MHz and ${\rm S/N}\geq3$, the 95\% Wilson intervals of the detection fractions are 43--90\% (non-V-shaped) and 16--75\% (V-shaped). The two classes show no significant difference, and J0923 and J1200 show that a V-shaped continuum can coexist with a radio detection.

\section{Spectral models for the J1047 turnover}
\label{app:j1047_models}

We compare three standard spectral forms with the radio spectrum of J1047 (Fig.~\ref{fig:j1047}), namely foreground free--free absorption (FFA), free--free absorption mixed with the synchrotron-emitting plasma, and synchrotron self-absorption (SSA). In all three models, the optically thin spectral index is fixed at $\alpha_{\rm thin}=-0.85$, as measured across the 4--8\,GHz band in 2018. The models are fitted by $\chi^2$ minimisation to the measured flux densities, with the non-detections included at their signed values. The fits do not include the variability uncertainty arising from the different epochs and angular resolutions of the data.

\paragraph{Homogeneous FFA.}
\label{app:j1047_hffa}
For a uniform ionised screen in front of the synchrotron source, with $\tff(\nu)=\tau_0(\nu/{\rm GHz})^{-2.1}$, where $\nu$ is observed frequency and $\tau_0$ is the optical depth at observed 1\,GHz,
\begin{equation}
S_\nu = S_0\left(\frac{\nu}{\rm GHz}\right)^{\alpha_{\rm thin}}\exp[-\tff(\nu)]\,.
\label{eq:hffa}
\end{equation}
The fit gives $\tau_0=3.3$. Converted to rest-frame frequency, Eq.~(\ref{eq:tff}) gives ${\rm EM}=\tau_0(1+z)^{2.1}/(3.28\times10^{-7})\simeq1.4\times10^7\,{\rm pc\,cm^{-6}}$ for $T_e=10^4$\,K. For $\Delta R=0.1$--$1$\,pc, $\ne=({\rm EM}/\Delta R_{\rm pc})^{1/2}\simeq4\times10^3$--$1.2\times10^4\,{\rm cm^{-3}}$, corresponding to $\Ne\simeq(4$--$11)\times10^{21}\,{\rm cm^{-2}}$. The model spectrum peaks at 2.7\,GHz in the observer frame. With $\alpha_{\rm thin}$ fixed, the peak lies between 0.8 and 5.8\,GHz at 95 per cent confidence (single-parameter profile likelihood with $\Delta\chi^2=3.84$ and the normalisation optimised), without allowance for variability or calibration errors.

\paragraph{Mixed FFA.}
\label{app:j1047_mffa}
If the synchrotron emissivity and the free--free absorbing gas are uniformly mixed along the line of sight, the attenuation factor becomes $(1-e^{-\tff})/\tff$ \citep{1991ApJ...378...65C}, giving
\begin{equation}
S_\nu = S_0\left(\frac{\nu}{\rm GHz}\right)^{\alpha_{\rm thin}}\frac{1-\exp[-\tff(\nu)]}{\tff(\nu)}\,.
\label{eq:mffa}
\end{equation}
The fit gives $\tau_0=7.5$, or rest-frame ${\rm EM}\simeq3.2\times10^7\,{\rm pc\,cm^{-6}}$ for $T_e=10^4$\,K. Over $\Delta R=0.1$--$1$\,pc, the corresponding column is $\Ne\simeq(5$--$17)\times10^{21}\,{\rm cm^{-2}}$. For absorber thicknesses in this range, which the spectrum itself does not constrain, both FFA geometries require an ionised column of order $10^{22}\,{\rm cm^{-2}}$, far below the $\sim10^{24}\,{\rm cm^{-2}}$ scattering column of the dense cocoon. A thinner absorber lowers the column further, so the emission measure is the robust quantity.

\paragraph{Synchrotron self-absorption.}
\label{app:j1047_ssa}
For a homogeneous synchrotron source with a power-law electron distribution \citep{1970ranp.book.....P},
\begin{equation}
S_\nu = S_p\left(\frac{\nu}{\nu_p}\right)^{5/2}
\frac{1-\exp\!\left[-\tau_p\left(\nu/\nu_p\right)^{\alpha_{\rm thin}-5/2}\right]}{1-\exp(-\tau_p)}\,,
\label{eq:ssa}
\end{equation}
where $\nu_p$ and $S_p$ are the frequency and flux density at the spectral peak. The SSA optical depth at the peak, $\tau_p$, satisfies
\begin{equation}
\frac{5}{2}\left(e^{\tau_p}-1\right)=\left(\frac{5}{2}-\alpha_{\rm thin}\right)\tau_p\,,
\end{equation}
which gives $\tau_p\simeq0.56$ for $\alpha_{\rm thin}=-0.85$. The fit gives $\nu_p\simeq2.7$\,GHz, and the source becomes optically thick ($\tau_{\rm SSA}>1$) below $\nu_p\,\tau_p^{1/(5/2-\alpha_{\rm thin})}\simeq2.3$\,GHz.

The fitted peak flux density, $S_p\simeq0.18$\,mJy, gives $T_b=1.22\times10^{12}(1+z)\,S_p({\rm Jy})\,\nu_{p,{\rm GHz}}^{-2}\,\theta_{\rm mas}^{-2}\simeq3.5\times10^9(\theta/0.1\,{\rm mas})^{-2}$\,K. 
% The SSA fit leaves $T_b$ to an independent constraint on size, magnetic field, or equipartition. 
The SSA fit alone does not determine $T_b$ without additional constraints on source size or magnetic field, or an equipartition assumption.
For a circular Gaussian source with negligible Doppler boosting, equipartition at $T_b\simeq5\times10^{10}$\,K would imply $\theta\simeq0.027$\,mas.

\end{appendix}

\end{document}